\documentclass[12pt]{article}

\usepackage{graphicx}

\begin{document}
\begin{center}

{\bf Born$-$Infeld-type modified gravity }\\
\vspace{5mm}
 S. I. Kruglov
 \footnote{serguei.krouglov@utoronto.ca}

\vspace{5mm}
\textit{Department of Physics, University of Toronto, \\60 St. Georges St.,
Toronto, ON M5S 1A7, Canada\\
Department of Chemical and Physical Sciences,\\ University of Toronto Mississauga,\\
3359 Mississauga Rd. N., Mississauga, ON L5L 1C6, Canada}
\end{center}

\begin{abstract}
We propose and investigate the modified Born$-$Infeld-type gravity model with the function
$F(R) = [1-(1-\beta R/\sigma)^\sigma]/\beta$.
At different values of the dimensionless parameter $\sigma$ the action is converted into some models including  general relativity ($\sigma=1$), the Starobinsky model ($\sigma=2$), the exponential model of gravity ($\sigma=\infty$), and a model proposed in \cite{Kruglov1}, \cite{Kruglov2} ($\sigma=0.5$). A bound on the parameters, $\beta(1-\sigma)/\sigma\leq 2\times 10^{-6}$ cm$^2$, are due to local tests. The constant curvature solutions are obtained, and we found the potential, and the mass of the scalaron in the Einstein frame. The model describes the inflation of the universe. The de Sitter spacetime is unstable and a Minkowsky spacetime is stable. We investigate the cosmological parameters and some critical points of autonomous equations.
\end{abstract}



\section{Introduction}

The inflation and the present time acceleration of the universe can be described by the modification of the general relativity (GR) making use of the replacement of the Ricci scalar $R\rightarrow F(R)$ in the Einstein$-$Hilbert action \cite{Faraoni}-\cite{Odintsov}. $F(R)$ gravity models modify GR and can be an alternative to $\Lambda$-Cold Dark Matter ($\Lambda$CDM) model describing the dark energy (DE) and the cosmic acceleration. The classical and quantum stabilities require the inequalities $F'(R)>0$, $F''(R)>0$ \cite{Appleby1} which do not define the function $F(R)$, and therefore, authors considered different functions $F(R)$.
Some examples of successful models of $F(R)$-gravity were studied in \cite{Hu}-\cite{Starobinsky}.

In this paper we propose the Born$-$Infeld (BI) type model with the function $F(R)$ containing two parameters $\beta$ and $\sigma$. For some values of $\sigma$ we come to known models: the Starobinsky model \cite{Starobinsky1}, exponential model of the gravity and the BI-like gravity model \cite{Kruglov1}, \cite{Kruglov2}.
Some BI-type gravity models were proposed in \cite{Deser}-\cite{Ketov}.

The singularities in BI electrodynamics, which are motivated by the string theory \cite{Fradkin}, are absent for point-like particles and the self-energy is finite \cite{Born}, \cite{Plebanski}. Therefore, it is interesting to investigate BI-type gravity. It should be mentioned that $F(R)$-gravity is the phenomenological model, that can describe the universe evolution, and may be an alternative to the $\Lambda$CDM model. Authors of the work \cite{Capozziello1} shown that for any $F(R)$-gravity model in Robertson$-$Walker space-time the stress-energy tensor can be represented in the form of a perfect fluid. This is important because in cosmology perfect fluids allow us to describe the Hubble flow from inflation to dark energy epochs. In \cite{Capozziello2} authors considered a cosmological model with a tachyon Dirac$-$Born$-$Infeld field interacting with canonical scalars. In the present paper we investigate $F(R)$-gravity model without tachyon fields.	

The structure of the paper is as follows. In Sec.2, we propose a new model of $F(R)$-gravity with the BI-type Lagrangian density with two parameters $\beta$ and $\sigma$. A bound on parameters $\beta$, $\sigma$, which follows from the laboratory experiment, was obtained. We found constant curvature solutions corresponding to de Sitter space-time in Sec. 3. In Sec. 4 the Einstein frame was considered corresponding to the scalar-tensor form of the model. We obtained the potential and the mass of the scalar degree of freedom. It was demonstrated that the de Sitter phase is unstable and the Minkowski space-time is stable. We calculated the slow-roll cosmological parameters of the model $\epsilon$ and $\eta$ in Sec. 5. In Sec. 6 some critical points of autonomous equations was studied. We evaluated the function $m(r)$ characterizing the deviation from the $\Lambda$CDM model. The results were discussed in Sec. 7.

\section{The Model}

We propose the modified Born$-$Infeld-type gravity model with the Lagrangian density
\begin{equation}
{\cal L}=\frac{1}{2\kappa^2}F(R)=\frac{1}{2\kappa^2}\frac{1}{\beta}\left[1-\left(1-\frac{\beta}{\sigma} R\right)^\sigma\right],
\label{1}
\end{equation}
where $\kappa=\sqrt{8\pi G}$, $G$ is the Newton constant, $\beta$ possesses the dimension of (length)$^2$ and $\sigma$ is dimensionless parameter.
The action for the pure gravitational field is given by
\begin{equation}
S=\int d^4x\sqrt{-g}{\cal L}=\int d^4x\sqrt{-g}\left[\frac{1}{2\kappa^2}F(R)\right].
\label{2}
\end{equation}
As particular cases at $\sigma=1$ we have the Einstein$-$Hilbert action, at $\sigma=2$ one comes to the Starobinsky model \cite{Starobinsky} that gives the self-consistent inflation \cite{Appleby}, at $\sigma=0.5$ - to the model proposed in \cite{Kruglov1}, \cite{Kruglov2}, and at $\sigma=\infty$ we come to the exponential model with the function
\begin{equation}
\lim_{\sigma\rightarrow\infty}F(R)=\frac{1}{\beta}\left[1-\exp(-\beta R)\right].
\label{3}
\end{equation}
Some exponential gravity-models were investigated in \cite{Zerbini}-\cite{Kruglov3}.
Thus, the Lagrangian (1) proposed gives a convenient parametrization that allows us to consider old models as well as to investigate new models for different parameters $\sigma$. We have the restriction $\beta R/\sigma\leq 1$ in order to have the real function (1).
The Taylor series for small values of $\beta R/\sigma$ give
\begin{equation}
F(R)= R+\frac{\beta(1-\sigma)}{2\sigma}R^2 +\frac{\beta^2(\sigma-2)(\sigma-1)}{6\sigma^2}R^3+... .
\label{4}
\end{equation}
Thus, at small values of $\beta R/\sigma$ we come to the Einstein$-$Hilbert action, i.e. the correspondence principle holds. We can obtain a bound on $\beta$ and $\sigma$, to passes local tests, with the help of the laboratory experiment \cite{Kapner}-\cite{Berry}, (see also \cite{Eingorn}), $F''(0)\leq 2\times 10^{-6}$ cm$^2$.  From Eq. (4) we find the bound:
\begin{equation}
\frac{\beta\left(1-\sigma\right)}{\sigma}\leq 2\times 10^{-6} \mbox{cm}^2.
\label{5}
\end{equation}

\section{Constant Curvature Solutions}

Field equations, when the Ricci scalar is a constant $R=R_0$, are given by \cite{Barrow}
\begin{equation}
2F(R_0)-R_0F'(R_0)=0.
\label{6}
\end{equation}
From Eqs. (1) and (6) we obtain
\begin{equation}
2\left[1-\left(1-\frac{\beta}{\sigma} R_0\right)^\sigma\right]=\beta R_0\left(1-\frac{\beta}{\sigma} R_0\right)^{\sigma-1}.
\label{7}
\end{equation}
The solution $R_0=0$ to Eq. (7) corresponds to the Minkowski spacetime.
Non-zero real numerical solutions to Eq. (7), for different values of the parameter $\sigma$, are given in Table 1.
\begin{table}[ht]
\caption{Approximate solutions to Eq. (7)}
\centering
\begin{tabular}{c c c c c c c c c c c c c c}
\hline
$\sigma$ & 0.4 & 0.5 & 0.6 & 0.7 & 0.8 & 0.9 & 3 & 4 & $\infty$ \\[0.5ex] 
\hline
$\beta R_0$ & 0.340 & 4/9 & 0.556 & 0.673 & 0.791 & 0.9 & -3$\sqrt{3}$ & -3.36 & -1.594\\
[1ex] 
\hline 
\end{tabular}
\label{table:Approx}
\end{table}
The classical stability condition $F'(R)>0$ and the quantum stability $F''(R)> 0$ \cite{Appleby} give the inequalities
\begin{equation}
F'(R)=\left(1-\frac{\beta}{\sigma} R\right)^{\sigma-1}>0,
\label{8}
\end{equation}
\begin{equation}
F''(R)=\frac{(1-\sigma)\beta}{\sigma}\left(1-\frac{\beta}{\sigma} R\right)^{\sigma-2}>0.
\label{9}
\end{equation}
Equation (8) leads to $\beta R/\sigma < 1$, and Eq. (9) gives $(1-\sigma)\beta/\sigma>0$ and $F(R)$ is a real function. As a result, at $0<\sigma<1$ the parameter $\beta$ has to be positive, $\beta>0$. But at $\sigma>1$ the parameter $\beta$ should be negative, $\beta<0$. In Table 1 for the case $\sigma>1$ we have the values $\beta<0$, $R_0>0$ leading to the quantum stability. Nontrivial solutions given in Table 1 correspond to the Schwarzschild$-$de Sitter spacetime and the trivial solution $R_0=0$ leads to the Minkowski spacetime.

The condition $F'(R_0)/F''(R_0)>R_0$ should be satisfied to describe DE which is future stable \cite{Schmidt}. Thus, from Eqs. (8) and (9) we come to
\begin{equation}
(2-\sigma)\beta R_0<\sigma.
\label{10}
\end{equation}
Nontrivial solutions given in Table 1 do not satisfy Eq. (10) and give unstable de Sitter spacetime describing the inflation. But the trivial solution $R_0=0$ obeys the inequality (10) and the flat spacetime is stable. The maximum of the effective potential in Einstein's frame corresponds to the constant curvature solutions. The model proposed can describe the inflation for the spacetime without matter.

\section{The Scalar-Tensor Form}

 The $F(R)$-model is represented in the Jordan frame and can be transformed into the Einstein frame corresponding to the scalar-tensor form. The metric may be transformed by the conformal transformation \cite{Magnano}
\begin{equation}
\widetilde{g}_{\mu\nu} =F'(R)g_{\mu\nu}=\left(1-\frac{\beta R}{\sigma} \right)^{\sigma-1}g_{\mu\nu}.
\label{11}
\end{equation}
Then one comes to the Lagrangian density corresponding to the scalar-tensor theory of the gravity
\begin{equation}
{\cal L}=\frac{1}{2\kappa^2}\widetilde{R}-\frac{1}{2}\widetilde{g}^{\mu\nu}
\nabla_\mu\phi\nabla_\nu\phi-V(\phi),
\label{12}
\end{equation}
where the scalar field $\phi$ and the potential $V(\phi)$ are given as follows:
\begin{equation}
\phi(R)=\frac{\sqrt{3}}{\sqrt{2}\kappa}\ln F'(R)=\frac{\sqrt{3}(\sigma-1)}{\sqrt{2}\kappa}\ln \left(1-\frac{\beta}{\sigma} R\right),
\label{13}
\end{equation}
\[
V(R)=\frac{RF'(R)-F(R)}{2\kappa^2F'^2(R)}
\]
\vspace{-7mm}
\begin{equation}
\label{14}
\end{equation}
\vspace{-7mm}
\[
=\frac{1}{2\kappa^2\beta}\left[\beta R\left(1-\frac{\beta}{\sigma}R\right)^{1-\sigma}- \left(1-\frac{\beta}{\sigma}R\right)^{2(1-\sigma)}+\left(1-\frac{\beta}{\sigma}R\right)^{2-\sigma}\right].
\]
In terms of the scalar field the potential (14) reads
\[
V(\phi)=\frac{1}{2\beta\kappa^2}\biggl\{\sigma\left[1-\exp\left(\frac{-\sqrt{2}\kappa\phi}{\sqrt{3}(1-\sigma)}\right)\right]
\exp\left(\frac{-\sqrt{2}\kappa\phi}{\sqrt{3}}\right)
\]
\begin{equation}
-\exp\left(\frac{-2\sqrt{2}\kappa\phi}{\sqrt{3}}\right)
\left[1-\exp\left(\frac{-\sqrt{2}\sigma\kappa\phi}{\sqrt{3}(1-\sigma)}\right)\right]\biggr\}.
\label{15}
\end{equation}
We obtain the potential for the exponential gravity at $\sigma\rightarrow \infty$ ($\beta<0$)
\begin{equation}
V(\phi)=\frac{1}{2|\beta|\kappa^2}\left[\left(\frac{\sqrt{2}}{3}\kappa\phi-1\right)
\exp\left(-\frac{\sqrt{2}}{3}\kappa\phi\right)+\exp\left(-\frac{2\sqrt{2}}{3}\kappa\phi\right)\right].
\label{16}
\end{equation}
The plots of the functions $2|\beta|\kappa^2V(\phi)$ vs. $\kappa \phi$ at some parameters $\sigma$ are depicted in Figs. 1 and 2.
\begin{figure}[h]
\includegraphics[height=4.0in,width=4.0in]{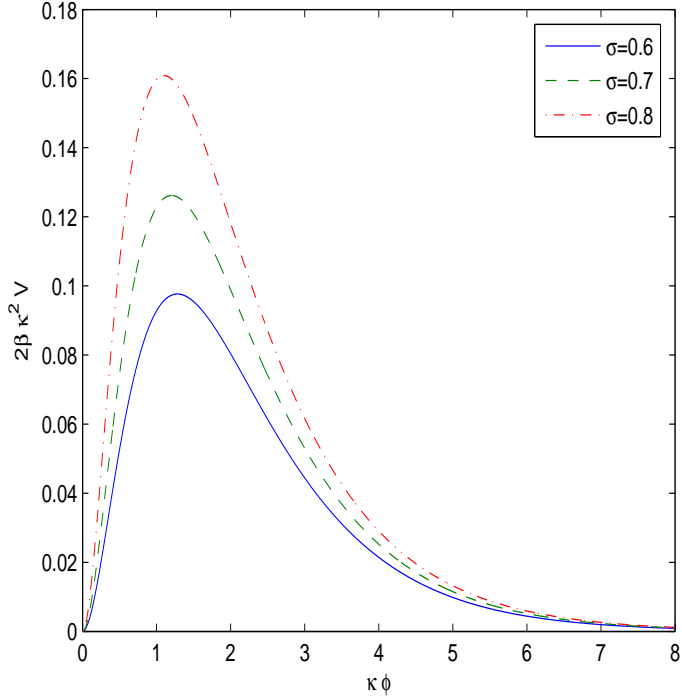}
\caption{\label{fig.1}The function $2\beta\kappa^2V$ vs. $\kappa\phi$ for $\sigma=0.6$, $0.7$, $0.8$}
\end{figure}
\begin{figure}[h]
\includegraphics[height=4.0in,width=4.0in]{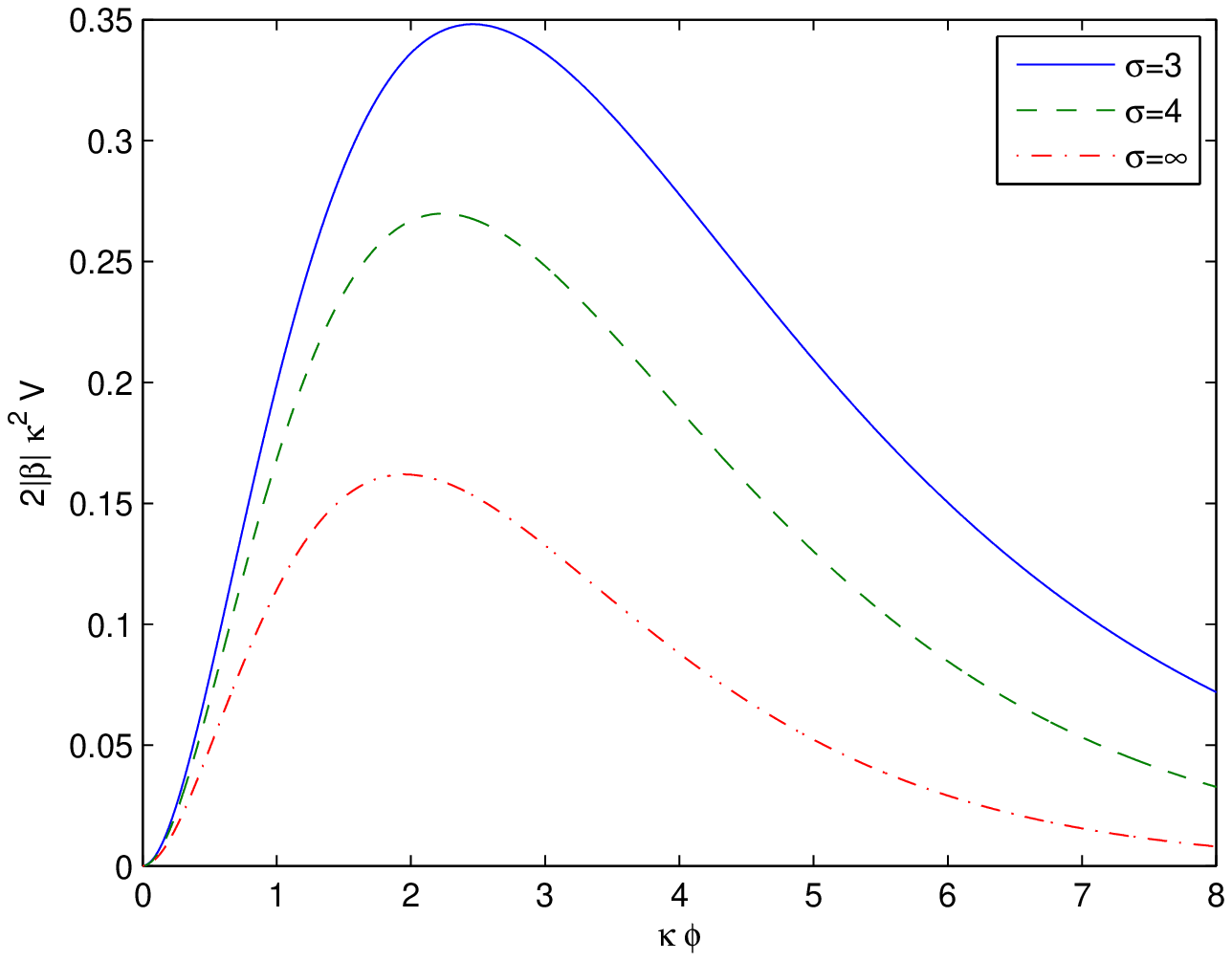}
\caption{\label{fig.2}The function $2|\beta|\kappa^2V$ vs. $\kappa\phi$ for $\sigma=3$, $4$, $\infty$}.
\end{figure}
The extremum of the potential is given by the equation
\begin{equation}
\frac{dV}{dR}=\frac{F''(R)\left[2F(R)-RF'(R)\right]}{2\kappa^2F^{'3}}=0.
\label{17}
\end{equation}
It follows from Eq. (17) that the constant curvature solutions correspond to the extremum of the potential.
The constant curvature solutions as functions of $\kappa\phi$ are represented in Table 2.
\begin{table}[ht]
\caption{Approximate solutions to Eq. (7)}
\centering
\begin{tabular}{c c c c c c c c c c c c c c}

\hline
$\sigma$ & 0.4 & 0.5 & 0.6 & 0.7 & 0.8 & 0.9 & 3 & 4 & $\infty$ \\[0.5ex]
\hline
$\kappa\phi$ & 1.398 & 1.346 & 1.280 & 1.196 & 1.099 & 0.259 & 2.462 & 2.240 & 1.952\\
[1ex]
\hline
\end{tabular}
\label{table:Approx}
\end{table}
The state of the flat spacetime ($R=0$) is stable and the states corresponding to Tables 1 and 2 are unstable.

We find that the mass squared of a scalaron is equal to
\[
m_\phi^2=\frac{d^2V}{d\phi^2} =\frac{1}{3}\left(\frac{1}{F''(R)}+\frac{R}{F'(R)}-\frac{4F(R)}{F^{'2}(R)}\right)
\]
\begin{equation}
=\frac{1}{3\beta}\left[\frac{4-3\sigma}{1-\sigma}\left(1-\frac{\beta}{\sigma}R\right)^{2-\sigma}+\beta R\left(1-\frac{\beta}{\sigma}R\right)^{1-\sigma}- 4\left(1-\frac{\beta}{\sigma}R\right)^{2(1-\sigma)}
\right].
\label{18}
\end{equation}
The plots of the functions $\beta m_\phi^2$ versus $\beta R$ are represented in Figs.3 and 4.
\begin{figure}[h]
\includegraphics[height=4.0in,width=4.0in]{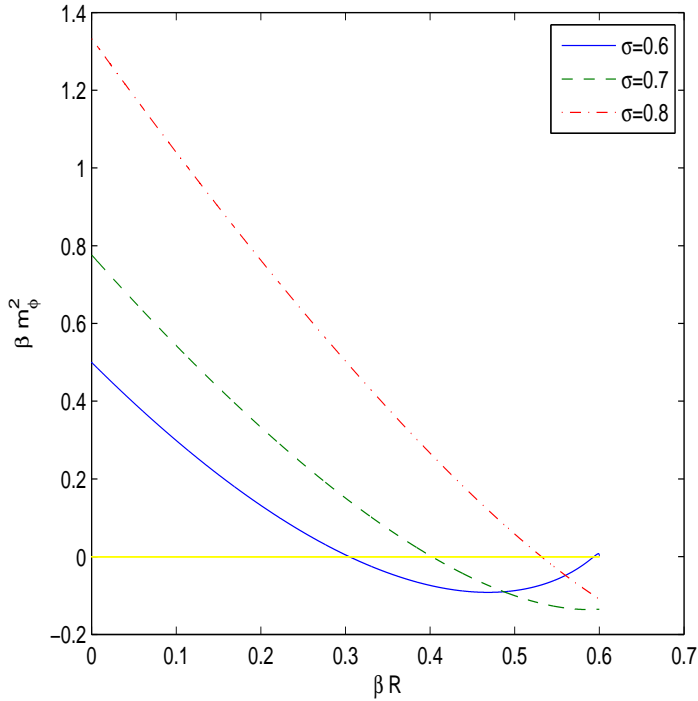}
\caption{\label{fig.3}The function $\beta m^2_\phi$ vs. $\beta R$ for $\sigma=0.6$, $0.7$, $0.8$.}
\end{figure}
\begin{figure}[h]
\includegraphics[height=4.0in,width=4.0in]{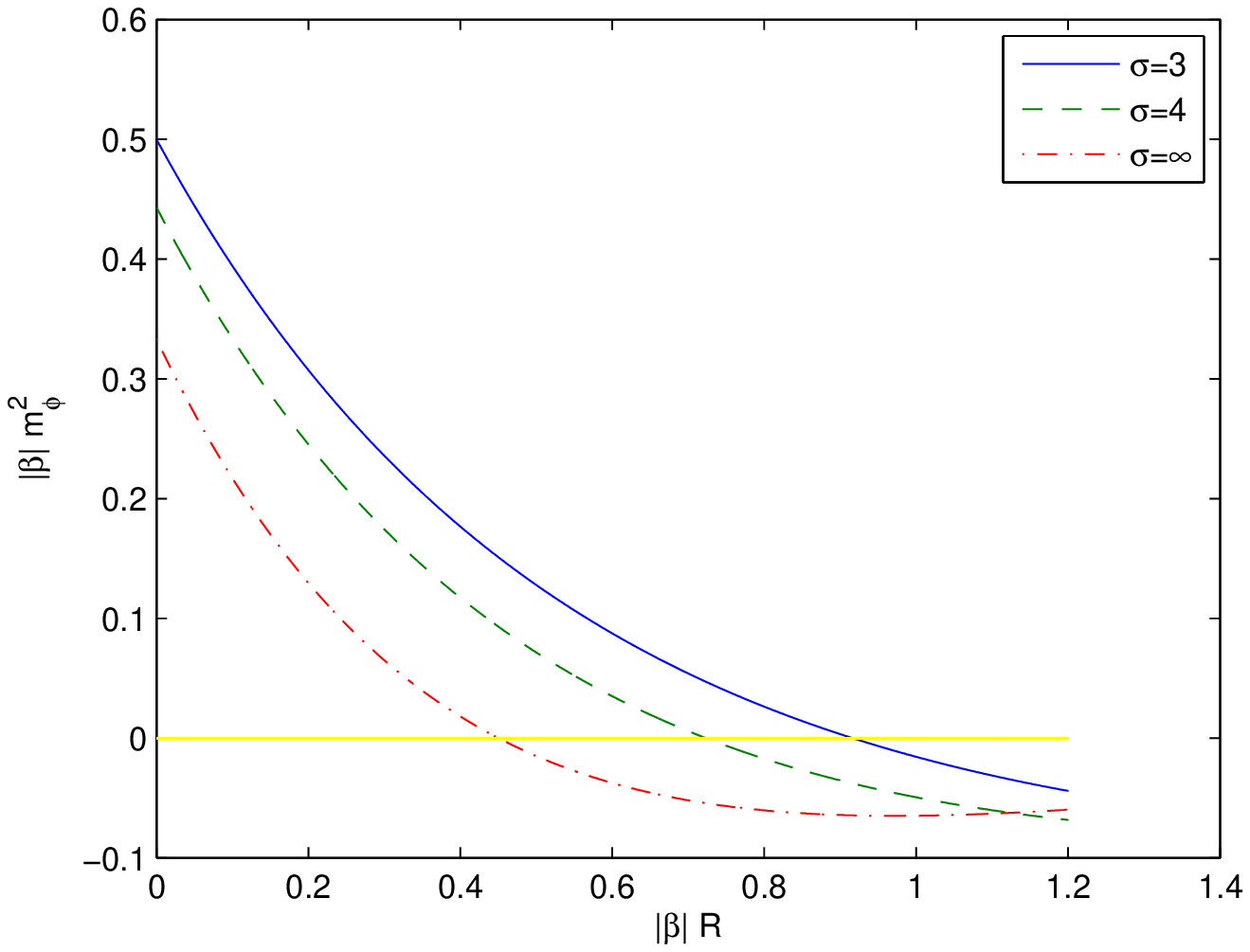}
\caption{\label{fig.4}The function $|\beta| m^2_\phi$ vs. $|\beta|R$ for $\sigma=3$, $4$, $\infty$.}
\end{figure}
It follows from numerical calculations and from Fig. 3 that $m_\phi^2$ is positive for $\sigma=0.6$ at $\beta R<0.3045$, for $\sigma=0.7$ at $\beta R<0.4022$, and for $\sigma=0.8$ at $\beta R<0.5317$.
We find from Fig. 4 and numerical calculations that $m_\phi^2$ is positive for $\sigma=3$ at $|\beta| R<0.916$, for $\sigma=4$ at $|\beta| R<0.720$, and for $\sigma=\infty$ at $|\beta| R<W_n(-4/e^2)+3\approx 0.450$.
Thus, the constant curvature solutions represented in Tables 1 and 2 give the negative values of $m^2_\phi$.  This tells us that states are non-stable \cite{Schmidt}.

\section{Cosmological Parameters}

Let us verify that corrections of to our gravity model are small comparing to GR for $R\gg R_0$, where $R_0$ is a curvature at the present time. This requirement gives \cite{Appleby1}
\begin{equation}
\mid F(R)-R\mid \ll R,~\mid F'(R)-1\mid \ll 1,~\mid RF''(R)\mid\ll 1.
\label{19}
\end{equation}
Making use of Eq. (4), we find that Eqs. (19) are satisfied for $|(1-\sigma)\beta R/\sigma|\ll 1$.\\
The slow-roll parameters are given by \cite{Liddle}
\begin{equation}
\varepsilon(\phi)=\frac{1}{2}M_{Pl}^2\left(\frac{V'(\phi)}{V(\phi)}\right)^2,~~~~
\eta(\phi)=M_{Pl}^2\frac{V''(\phi)}{V(\phi)}.
\label{20}
\end{equation}
The reduced Planck mass is $M_{Pl}=\kappa^{-1}$. From Eq. (14), one finds the slow-roll parameters as follows:
\begin{equation}
\varepsilon=\frac{1}{3}\left[\frac{RF'(R)-2F(R)}{RF'(R)-F(R)}\right]^2=\frac{1}{3}\left[\frac{x(1-x/\sigma)^{\sigma-1}
-2+2(1-x/\sigma)^\sigma}{x(1-x/\sigma)^{\sigma-1}-1+(1-x/\sigma)^\sigma}\right]^2,
\label{21}
\end{equation}
\[
\eta=\frac{2}{3}\left[\frac{F^{'2}(R)+F''(R)\left[RF'(R)-4F(R)\right]}{F''(R)\left[RF'(R)-F(R)\right]}\right]
\]
\begin{equation}
=\frac{2\left[\frac{4-3\sigma}{1-\sigma}(1-x/\sigma)^{2\sigma-2}-4(1-x/\sigma)^{\sigma-2}+x(1-x/\sigma)^{2\sigma-3}\right]}
{3\left[x(1-x/\sigma)^{2\sigma-3}-(1-x/\sigma)^{\sigma-2}+(1-x/\sigma)^{2\sigma-2}\right]},
\label{22}
\end{equation}
where $x=\beta R$. The slow-roll approximation takes place if the conditions $\epsilon\ll 1$, $\mid\eta\mid\ll 1$ hold. The plots of the functions $\epsilon$, $\eta$ are depicted in Figs. 5 - 8.
\begin{figure}[h]
\includegraphics[height=4.0in,width=4.0in]{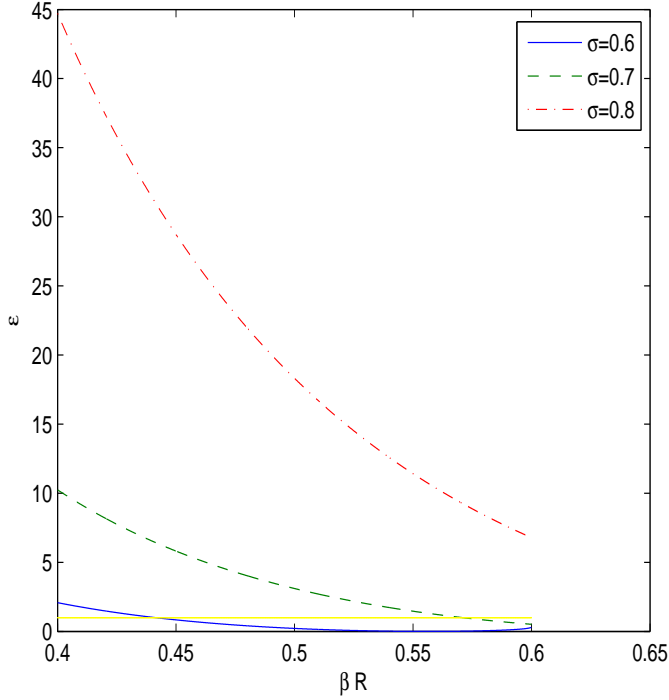}
\caption{\label{fig.5} The function $\varepsilon$ vs. $\beta R$ for $\sigma=0.6$, $0.7$, $0.8$. }
\end{figure}
\begin{figure}[h]
\includegraphics[height=4.0in,width=4.0in]{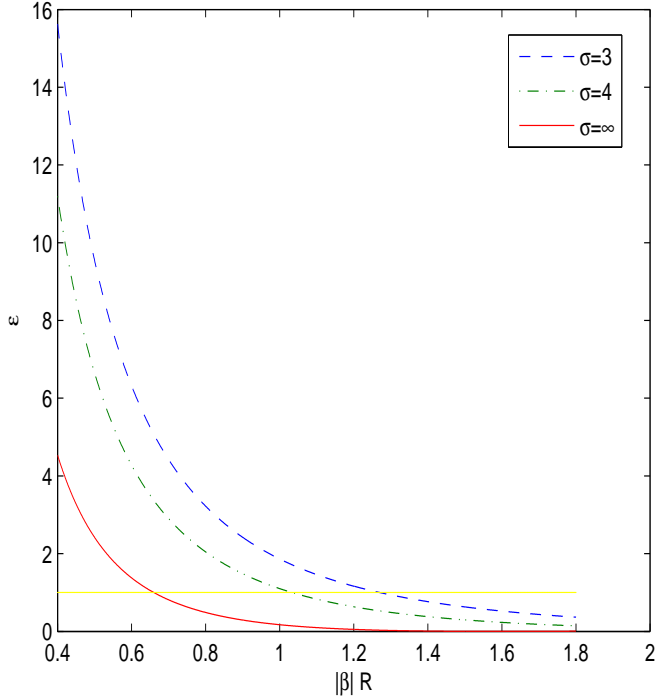}
\caption{\label{fig.6}The function $\varepsilon$ vs. $|\beta|R$ for $\sigma=3$, $\sigma=4$, $\sigma=\infty$. }
\end{figure}
\begin{figure}[h]
\includegraphics[height=4.0in,width=4.0in]{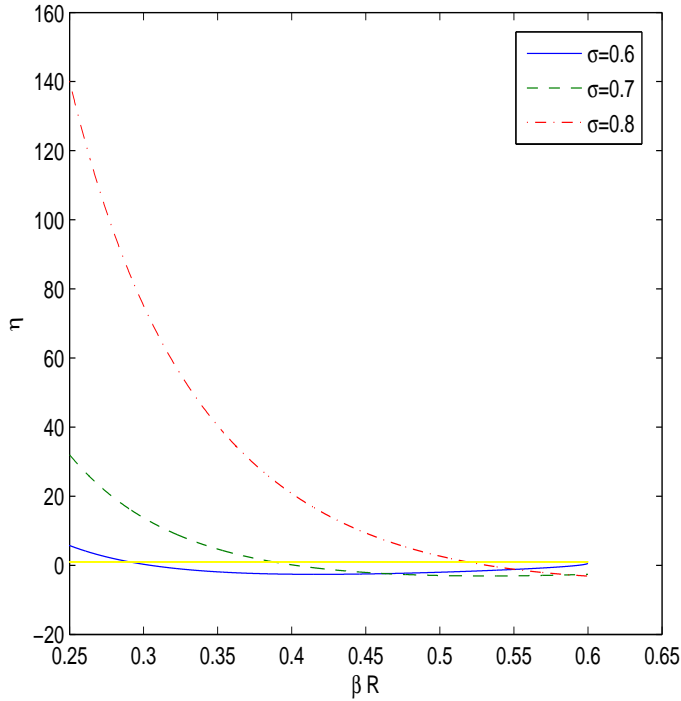}
\caption{\label{fig.7} The function $\eta$ vs. $\beta R$ for $\sigma=0.6$, $0.7$, $0.8$. }
\end{figure}
\begin{figure}[h]
\includegraphics[height=4.0in,width=4.0in]{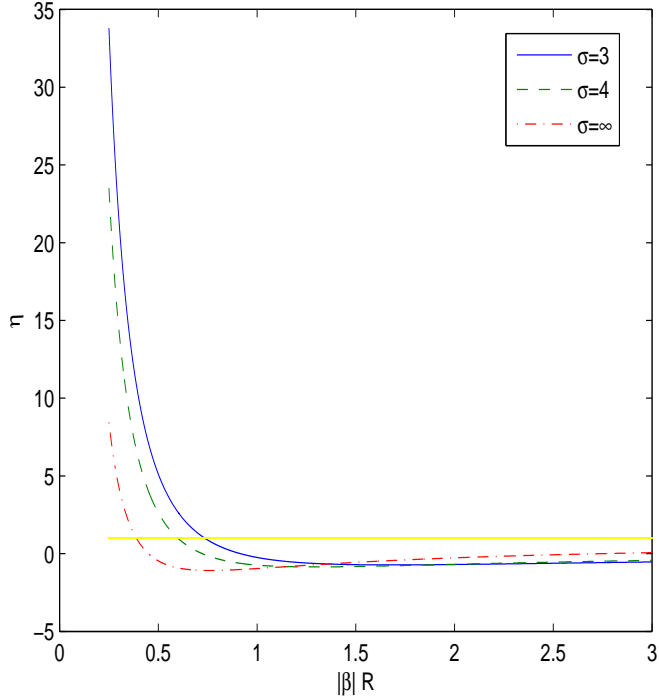}
\caption{\label{fig.8}The function $\eta$ vs. $|\beta|R$ for $\sigma=3$, $\sigma=4$, $\sigma=\infty$. }
\end{figure}
In Tables  3 and 4 the approximate solutions of equations $\varepsilon=1$ and $|\eta|=1$ are given.
\begin{table}[ht]
\caption{Approximate solutions to equation $\varepsilon=1$}
\centering
\begin{tabular}{c c c c c c c c c c c c c c}
\hline
$\sigma$ & 0.5 & 0.6 & 0.7 & 0.8 & 3 & 4 & $\infty$ \\[0.5ex]
\hline
$|\beta|R$ & 0.333 & 0.441 & 0.571 & 0.724 & 1.270 & 1.033 & 0.660\\
[1ex]
\hline
\end{tabular}
\label{table:Approx}
\end{table}
\begin{table}[ht]
\caption{Approximate solutions to equation $|\eta|=1$}
\centering
\begin{tabular}{c c c c c c c c c c c c c c}
\hline
$\sigma$ & 0.5 & 0.6 & 0.7 & 0.8 & 3 & 4 & $\infty$ \\[0.5ex]
\hline

$|\beta|R$ & 0.215 & 0.290 & 0.388 & 0.518 & 0.735 & 0.599 & 0.389, 0.944, 0.640 \\
[1ex]
\hline
\end{tabular}
\label{table:Approx}
\end{table}
\\
These values, given in Tables 3 and 4, can be used for calculations of the age of the inflation.
One can evaluate the age of the inflation by calculating the $e$-fold number \cite{Liddle}
\begin{equation}
N_e\approx \frac{1}{M_{Pl}^2}\int_{\phi_{end}}^{\phi}\frac{V(\phi)}{V'(\phi)}d\phi,
\label{23}
\end{equation}
where the $\phi_{end}$ corresponds to the end of the inflation. From Eqs. (13) and (14) we obtain the number of $e$-foldings
\begin{equation}
N_e\approx \frac{3(1-\sigma)}{2\sigma}\int_{x_{end}}^{x_{start}}\frac{x(1-x/\sigma)^{\sigma-1}-1+(1-x/\sigma)^{\sigma}}
{(1-x/\sigma)[x(1-x/\sigma)^{\sigma-1}-2+2(1-x/\sigma)^{\sigma}]}dx,
\label{24}
\end{equation}
were $x_{end}=\beta R_{end}$ corresponds to the end of the inflation when $\epsilon$ or $|\eta |$ are close to $1$.
At $\sigma\rightarrow\infty$ we obtain from Eq. (24)
\begin{equation}
N_e\approx \frac{3}{2}\int_{x_{end}}^{x_{start}}\frac{1+x-\exp(x)}{x+2-2\exp(x)}dx.
\label{25}
\end{equation}
Remember that for $\sigma>1$ the parameter $\beta$ is negative, so $x=\beta R<0$.
For example, at $|x_{end}|=0.94$ (see Tables 3 and 4) and $|x_{start}|=82$ we obtain from Eq. (25) $N_e\approx 60$, i.e. reasonable age of the inflation \cite{Liddle}.
As a result, the model under consideration describes the inflation and reproduces the necessary age of the inflation for some parameter $\sigma$.

\section{Critical Points and Stability}

To investigate critical points and stability, for FRW cosmology, we explore the dimensionless parameters \cite{Amendola}
\begin{equation}
x_1=-\frac{\dot{F}'(R)}{HF'(R)},~x_2=-\frac{F(R)}{6F'(R)H^2},~x_3=\frac{\dot{H}}{H^2}+2,
\label{26}
\end{equation}
\begin{equation}
m=\frac{RF''(R)}{F'(R)},~~~~r=-\frac{RF'(R)}{F(R)}=\frac{x_3}{x_2},
\label{27}
\end{equation}
where $H=\dot{a}/a$ is a Hubble parameter and the dot denotes the derivative with respect to the cosmic time. The $m(r)$   describes the deviation from the $\Lambda$CDM model. Making use of variables (26) equations of motion, in the absence of the radiation, $\rho_{\mbox{rad}}=0$ ($x_4=0$), can be written in the form of autonomous equations \cite{Amendola}:
\begin{equation}
\frac{dx_i}{dN}=f_i(x_1,x_2,x_3)~~~~(i=1,2,3),
\label{28}
\end{equation}
were $N\equiv N_e=\ln a$ is the number of $e$-foldings. The functions $f_i(x_1,x_2,x_3)$ are as follows:
\[
f_1(x_1,x_2,x_3)=-1-x_3-3x_2+x_1^2-x_1x_3,
\]
\begin{equation}
f_2(x_1,x_2,x_3)=\frac{x_1x_3}{m}-x_2\left(2x_3-4-x_1\right),
\label{29}
\end{equation}
\[
f_3(x_1,x_2,x_3)=-\frac{x_1x_3}{m}-2x_3\left(x_3-2\right).
\]
One can investigate the critical points of the equation system by studying the function $m(r)$. From Eqs. (1), (8), (9) and (27), we obtain
\[
m=\frac{x(1-\sigma)}{\sigma (1-x/\sigma)},
\]
\vspace{-7mm}
\begin{equation}
\label{30}
\end{equation}
\vspace{-7mm}
\[
r=\frac{x}{1-x/\sigma -(1-x/\sigma)^{1-\sigma}},
\]
where $x=\beta R$. The plots of the function $m(r)$ is depicted in Figs. 9 and 10.
\begin{figure}[h]
\includegraphics[height=4.0in,width=4.0in]{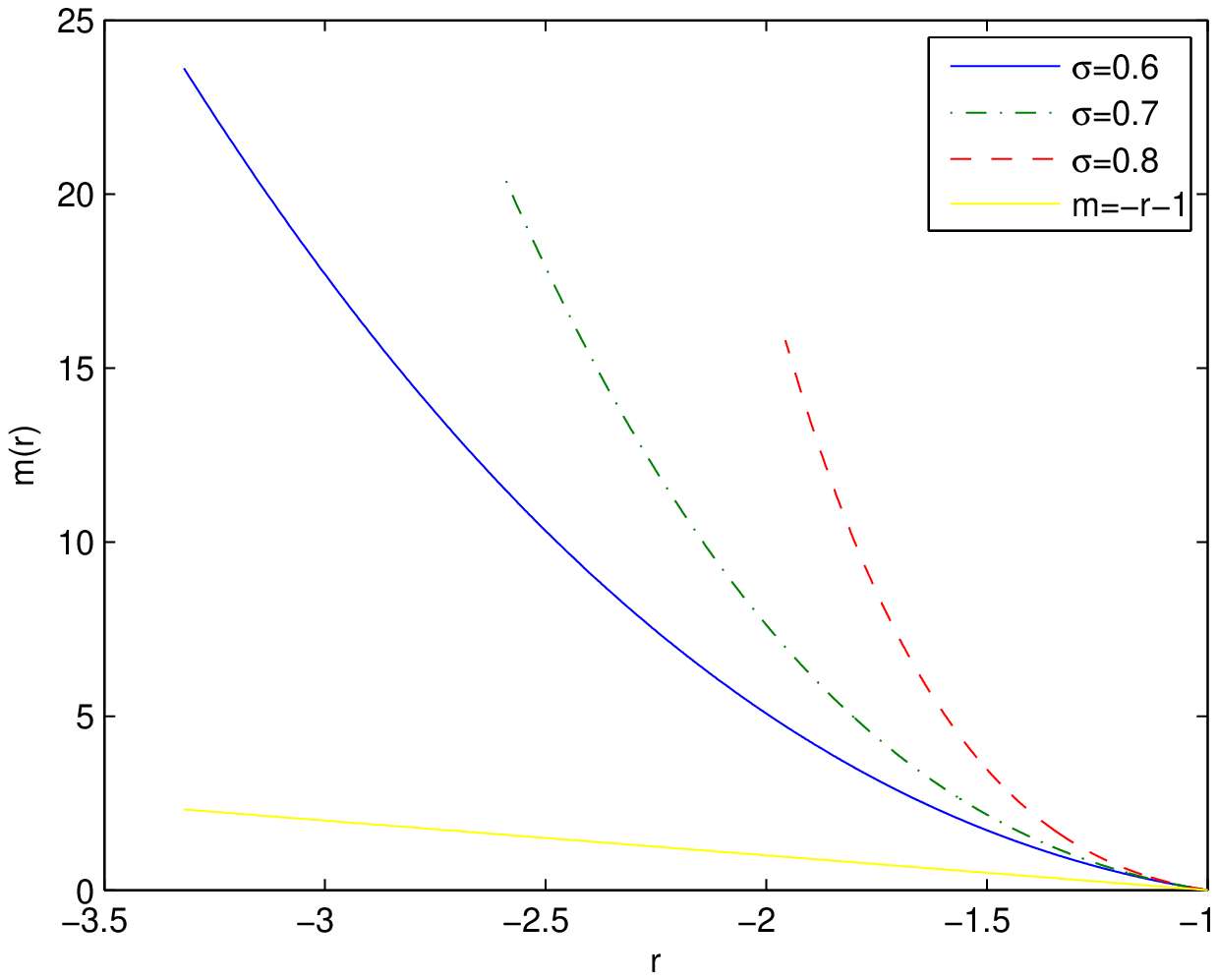}
\caption{\label{fig.9} The function $m(r)$ for $\sigma=0.6$, $0.7$, $0.8$). }
\end{figure}
\begin{figure}[h]
\includegraphics[height=4.0in,width=4.0in]{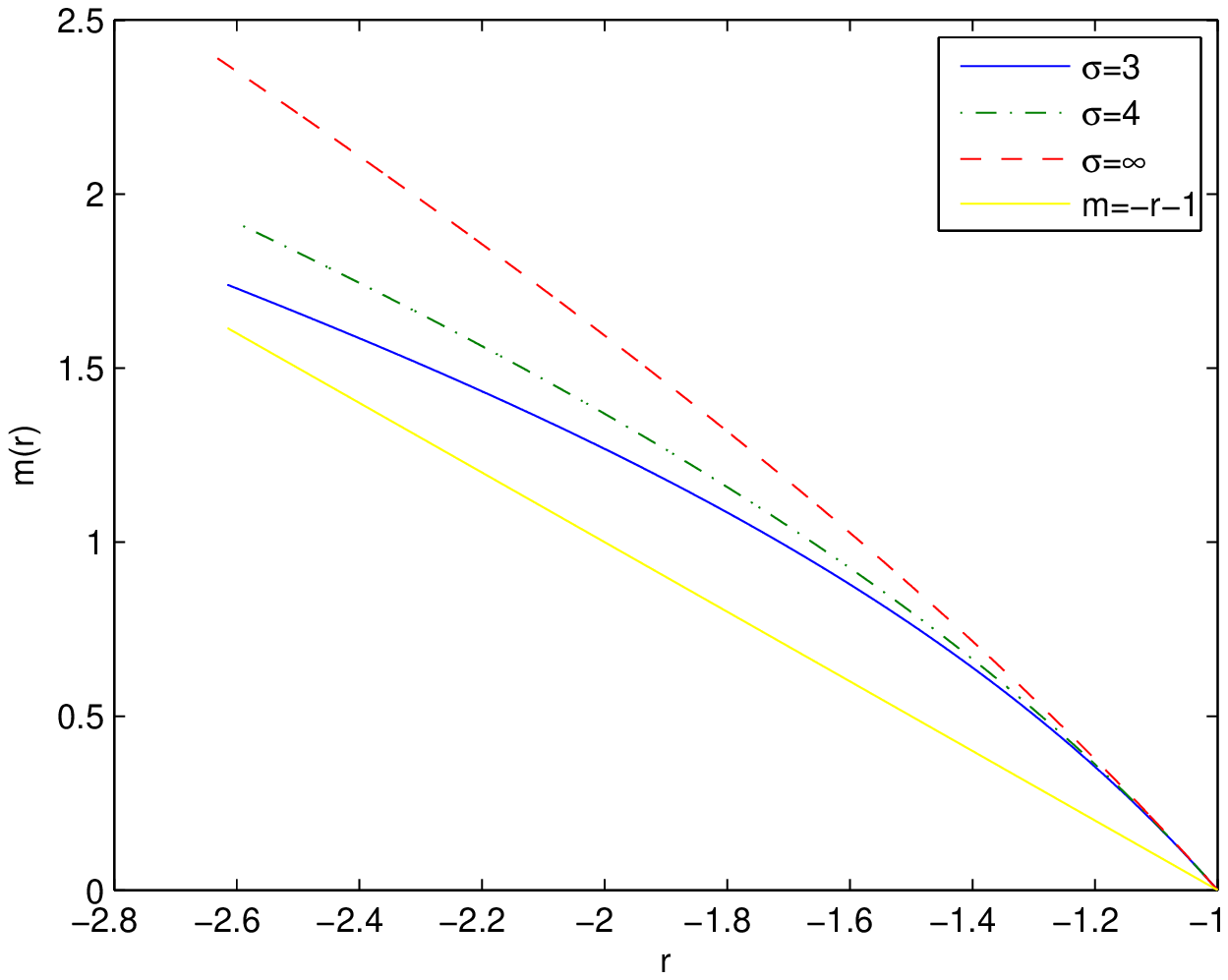}
\caption{\label{fig.10} The function $m(r)$ for $\sigma=3$, $4$, $\infty$). }
\end{figure}
In the absence of radiation, $x_4 = 0$, the de Sitter point $P_1$ \cite{Amendola}  corresponds to the parameters $x_1=0$, $x_2=-1$, $x_3=2$. Making use of Eqs. (6), and (27), this point corresponds to the constant curvature solutions ($\dot{H}=0$) and that mimics the cosmological constant $\Lambda=R_0/4$ ($H=\sqrt{\Lambda/3}$).
For this point the matter energy fraction parameter gives $\Omega_{\mbox{m}}=1-x_1-x_2-x_3=0$. The effective equation of state (EoS) parameter $w_{\mbox{eff}}=p/\rho=-1-2\dot{H}/(3H^2)=-1$ ($p$ is the pressure and $\rho$ is the energy density) and this corresponds to DE. According to Figs. 8 and (9) $1< m(r=-2)$ and the constant curvature solutions give unstable the de Sitter spacetime \cite{Amendola}. Prior to late-time acceleration, a viable matter dominated epoch exists for the critical point $P_5$ with EoS of a matter era $w_{\mbox{eff}}=0$, $x_3=1/2$, $a=a_0t^{2/3}$ with $m\approx 0$, $r\approx -1$ \cite{Amendola}. According to Eqs. (30) this condition is satisfied (see also Figs. (9) and (10). The point $P_5$ belongs to the equation $m=-r-1$ with the solution $m=0$, $r=-1$. Making use of Eq. (30) we find that $r=-1$ corresponds to $x=0$ ($R=0$). The standard matter era exists if the condition $m'(r=-1)>-1$ holds \cite{Amendola}. We obtain the derivative $m'(r)=(dm/dx)(dx/dr)$ from Eqs. (30)
\begin{equation}
\frac{dm}{dr}=\frac{(1-\sigma)[(1-x/\sigma)^\sigma-1]^2}
{\sigma(1-x/\sigma)^\sigma[(1-x/\sigma)^\sigma+x-1]}.
\label{31}
\end{equation}
For the case $\sigma\rightarrow\infty$ ($\beta<0$, $x=-|x|$) we find from Eq. (31)
\begin{equation}
\frac{dm}{dr}=-\frac{(\exp(-|x|)-1)^2}
{1-(|x|+1)\exp(-|x|)}~~~~(\sigma\rightarrow\infty).
\label{32}
\end{equation}
It follows from Eqs. (31) and (32) that $\lim_{x\rightarrow 0}m'(r)=-2$. Thus, the condition $m'(r=-1)>-1$, to have a standard matter dominated epoch, is violated.
This model describes the inflation but not the current acceleration. One can modify the model introducing the cosmological constant by the replacement $F(R)\rightarrow F(R)-\Lambda$ to have the current universe acceleration.
For a detailed  description of the universe evolution one has to solve the autonomous equations \cite{Amendola}.

\section{Conclusion}

We have proposed a new $F(R)$ gravity model with two scales ($\beta$ and $\sigma$). As particular cases this model includes the Starobinsky model ($\sigma=2$), the BI-like model \cite{Kruglov1}, \cite{Kruglov2} ($\sigma=0.5$) and the exponential model ($\sigma=\infty$). This model describes inflation with the de Sitter solution. From the local tests we found the bound on the constants $\beta$ and $\sigma$, Eq. (5). It was demonstrated that the de Sitter spacetime is unstable and the solution with zero curvature is stable. At small $R$ the action becomes the EH action for any parameter $\sigma$. This model with new gravitational physics describes DE dynamically. We considered the scalar-tensor form of the model corresponding to the Einstein frame. The potential and the mass of the scalaron (the effective scalar field) were found. We have obtained ranges where the slow-roll parameters of $\epsilon$, $\eta$ are small. Some critical points of autonomous equations have been considered and
the function m(r) characterizing the deviation from the $\Lambda$CDM-model was evaluated. It was demonstrated that the necessary conditions for the standard matter era are not satisfied. Therefore, this model describes early-time inflation but does not give the complete description all stages of the universe evolution. But to explain the current acceleration of the universe one can introduce the cosmological constant.

\end{document}